\documentclass[apj]{emulateapj}
\usepackage{natbib}
\usepackage{amssymb}
\usepackage{graphicx}
\usepackage{subfigure}
\usepackage{bm}

\newcommand{\etal}{et~al.\@}
\newcommand{\eg}{e.g.\@}
\newcommand{\ie}{i.e.\@}

\newcommand{\ergs}{erg s$^{-1}$ }
\def\arcsec{\hbox{$^{\prime\prime}$}}
\def\arcseck{\hbox{$^{\prime\prime}$ }}

\newcommand{\RXTEk}{{\it RXTE }}
\newcommand{\RXTE}{{\it RXTE}}
\newcommand{\Swiftk}{{\it Swift} }
\newcommand{\Swift}{{\it Swift}}
\newcommand{\Chandrak}{{\it Chandra }}
\newcommand{\Chandra}{{\it Chandra}}

\shorttitle{Does the 62-day X-ray periodicity come from ULX M82 X-1? }

\begin{document}

\title{Does the 62-day X-ray periodicity come from ULX M82 X-1? }


\author{Yanli Qiu\altaffilmark{1,}\altaffilmark{2},
JiFeng Liu\footnotemark[$\dagger$]\altaffilmark{1},
Jincheng Guo\altaffilmark{1,}\altaffilmark{2}
and Jing Wang\altaffilmark{3}}
\altaffiltext{1}{Key Laboratory of Optical Astronomy, National Astronomical Observatories, Chinese Academy of Sciences, 20A Datun Road, Chaoyang Distict, Beijing 100012, China}
\altaffiltext{2} {University of Chinese academy of sciences, No.19A Yuquan Road, Beijing 100049, China}
\altaffiltext{3}{Key Laboratory of Space Astronomy and Technology, National Astronomical Observatories, Chinese Academy of Sciences, Beijing 100012, China}

\footnotetext[$\dagger$]{Send correspondence to jfliu@nao.cas.cn}

\begin{abstract}
M82 X-1 is the brightest ultraluminous X-ray source in starburst galaxy M82 and is one of the best intermediate mass black hole candidates. Previous studies based on the {\it Rossi X-ray Timing Explorer }/Proportional Counter Array
(\RXTE/PCA) reported a regular X-ray flux modulation of M82 with a period of 62 days, and attributed this periodic modulation to M82 X-1. However, this modulation is not necessarily from M82 X-1 because \RXTE/PCA has a very poor
spatial resolution of $\sim1^\circ$. In this work, we analyzed 1000 days of monitoring data of M82 X-1 from the {\it Swift}/X-ray telescope (XRT), which has a much better spatial resolution than \RXTE/PCA. The periodicity distribution map of M82 reveals that the 62-day periodicity is most likely not from M82 X-1, but from the summed contributions of several periodic X-ray sources 4\arcseck southeast of M82 X-1.  However, \Swift/XRT is not able to resolve those periodic sources and locate the precise origin of the periodicity of M82.  Thus, more long-term observations with higher spatial resolution are required.
\end{abstract}


\section{Introduction}\label{sec:intro}

Ultraluminous X-ray sources (ULXs; Fabbiano 2006) discovered in other galaxies
exhibit high X-ray luminosities $L_X\ge2\times10^{39}$ erg $s^{-1}$, and are possibly
the long sought after intermediate mass black holes (IMBHs) radiating at sub-Eddinton
levels.
Recent studies have shown that, except for a few cases, many ULXs can be
stellar mass black holes in an ultraluminous state under supercritical
accretion \citep{2009MNRAS.397..124G}.
Indeed, optical monitoring campaigns of two ULXs, M101 ULX-1
\citep{2013Natur.503..500L} and NGC 7793 P13 \citep{2014Natur.514..198M}, have
confirmed both to be stellar mass black holes, while {\it NuSTAR} observations of M82
have discovered that a transient ULX, M82 X-2, is actually powered by an
accreting neutron star \citep{2014Natur.514..202B}.

M82 X-1 (CXO J095550.2+694047) is the brightest ULX in the starburst galaxy, M82,
at a distance of 3.6Mpc \citep{1994ApJ...427..628F}, and arguably the best
IMBH candidate in the local universe.
M82 X-1 exhibits extremely high luminosities that can reach $10^{41}$ $erg\ s^{-1}$ in
the $0.5-10$ keV bands \citep{2001MNRAS.321L..29K, 2001ApJ...547L..25M,
2009ApJ...692..653K}, suggesting an IMBH of about $10^3 M_\odot$ if radiating at the
Eddington level.
Recently, Pasham et al. (2014) re-analyzed 6-year {\it Rossi X-ray Timing
Explorer }(\RXTE) X-ray observations and revealed the high-frequency, 3:2
ratio, twin-peak quasi-periodic oscillations (QPOs) of 3.3 Hz and 5Hz, which, in combination with the
low-frequency QPO revealed by an {\it XMM-Newton} observation, suggests an IMBH of
$415\pm63M_\odot$ under the relativistic precession model
\citep{2014MNRAS.437.2554M}.

Interestingly, X-ray observations of M82 with the Proportional Counter Array
(PCA) on board  \RXTEk over eight months has revealed regular modulations of about
20\% with a period of about 62 days (\citealt{2006ApJ...646..174K,
2007ApJ...669..106K}, hereafter KF07).
\citet{2006ApJ...646..174K} interpreted such a period as the orbital period for
M82 X-1, and consequently derived a giant or supergiant companion for M82 X-1.
However, the analysis of another 4.5-year \RXTE/PCA light curve of M82 has
shown a sudden phase shift of the 62-day modulation, suggesting the 62-day
X-ray period may be caused by a precessing accretion disk around the black hole
(\citealt{2013ApJ...774L..16P}, hereafter PS13).

While previous studies have attributed the periodic modulations in the
\RXTE/PCA light curves to \mbox{M82 X-1}, the brightest X-ray source in M82, it
is not necessarily true because \RXTE/PCA has a very poor spatial resolution of
$\sim1^\circ$, and the flux includes contributions from all X-ray sources in
M82 ($11^\prime \times 4^\prime$) and nearby galaxies.
In contrast, the X-ray telescope (XRT) on board the \Swiftk satellite has a much
better spatial resolution with a half power diameter of 18\arcsec, and may
resolve the bright point sources in M82.  In this Letter, we use \Swift/XRT
observations over 1000 days to scrutinize the X-ray periodicity of M82.
The data analysis and results are presented in \S 2, and the discussion follows
in \S3.

\section{DATA ANALYSIS AND RESULTS}

The \Swift/XRT observations of M82 operated in the photon-counting mode are
used in this work, which are retrieved from the HEASARC archive.
These include a total of 106 observations spanning from 2012 April to 2015
January. 70 observations in $2014$ are not used because they are badly
contaminated by the supernova SN 2014J in M82.
Inspection of these XRT observations confirms that X-ray photons from M82 X-1
and nearby sources dominate the field within $\sim30^\prime$ of M82.

\subsection{Stacked images of M82 with \rm{Swift} observations}

There are three other ULXs near M82 X-1, \eg, M82 X-2, CXOU J095551.2 +694044
(hereafter $X-3$) and CXOU J095550.6 +694944 (hereafter X-4; \citealt{2005A&A...429.1125L, 2007ApJ...671..349K, 2003ApJ...586L..61S}) with
$2\arcsec-6\arcsec$ separations from each other.
Those four ULXs can only be clearly resolved by the {\it Chandra X-ray observatory}
(see Figure 1(a)), unfortunately there are not enough \Chandrak observations of
M82 X-1 in order to perform periodicity analysis.
In contrast, \Swift/XRT has more than a hundred observations of M82 X-1, and it
can marginally resolve the four ULXs in a single observation (Figure 1(b)),
though the claimed half power diameter of the point spread function of XRT is
18\arcsec.
The image spatial resolution of XRT is one pixel (2\arcsec.36), but the
positional accuracy from centroiding can be higher when there are ample
photons.
To reveal of all the faint and bright X-ray sources with more accurate positions,
we stacked the total 106 XRT observations. Before image stacking, one pixel was
equally divided into $10\times10$ parts, and each part (a sub-pixel) was
assigned 1\% of the pixel counts.  Then the total 106 images in sub-pixels were
matched under WCS coordinates and stacked into one (Figure 1(c)).

This stacked image (Figure 1(c)) reveals that the four ULXs are not blurred
into one unresolved source, but an asymmetric structure in which X-1 and the
other three ULXs are separated.
X-2 is the second brightest X-ray source in M82, but it is not evident in
Figure 1(c) because it is a transient and, for most of the time, it is in a quiescent
state.  Among the 106  \Swiftk observations, only a dozen observations in which
X-2 is brighter than X-3 and X-4.  X-3 is also a variable X-ray source, and is
visible in 46 observations.  X-4 was reported as a  supernova remnant
candidate \citep{2007ApJ...671..349K, 2011MNRAS.414.1329C}, which is supposed
to have relatively constant flux.

Furthermore, to inspect whether XRT can resolve X-1 from X-3 or not, 46
observations in which X-3 is brighter than the nearby background were selected
, and stacked into one image (Figure 1(d)). The rest of the 60 observations are
stacked and shown in Figure 1(e).
These figures imply that X-1 can be roughly resolved from  X-3,
while counts from X-2, X-3, and X-4 within 4\arcseck mix up with each other and
can hardly be resolved by XRT.  Thus, sources with separations larger than
4\arcseck can marginally resolved by XRT, but not vice versa.

\begin{figure*}
\center
{\includegraphics[width=5in]{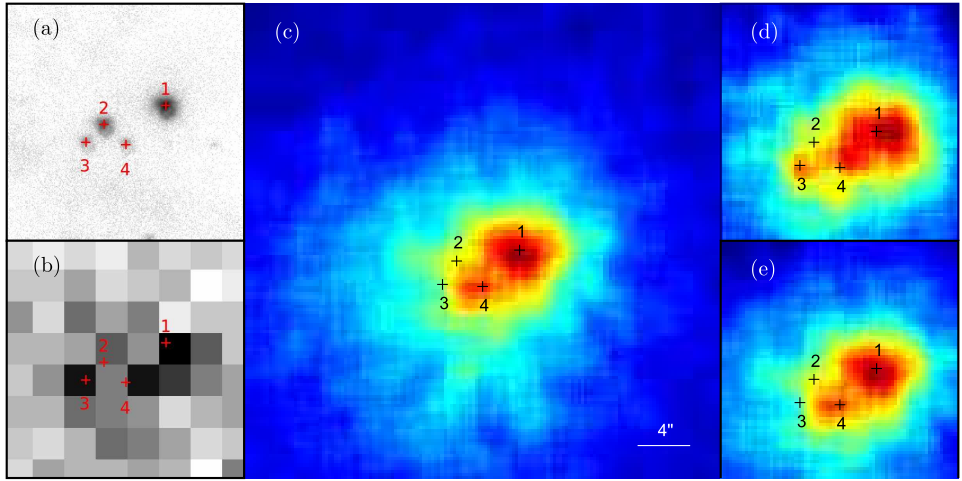}}
\label{f1}
\caption{
X-ray images of M82 from \Chandra/ACIS and \Swift/XRT. Each panel is shown on
the same spatial scale, and north is up.  The spatial size of (a), (b), (d), and
(e) is $18\arcsec\times18\arcsec$, and the size of (c) is
$36\arcsec\times36\arcsec$.  The positions of X-1, X-2, X-3, and X-4 are marked
by crosses and corresponding numbers.  (a) \Chandra/ACIS image from
observation ObsID 5644.  (b) \Swift/XRT image from observation ObsID 00032503069, which shows that the four
ULXs can be marginally resolved by  \Swift/XRT.  (c) The stacked image of all
106 \Swift/XRT observations used in this study. The counts  concentrated on X-1 and
X-4 and reveal an asymmetric structure.  (d) The stacked image of 46
\Swift/XRT observations in which X-3 is obviously brighter than the background.
(e) The stacked image of the remaining 60 observations.
}

\end{figure*}

\subsection{Phase offset of the 62 day X-ray Period}

\begin{figure*}[!htb]
\plottwo{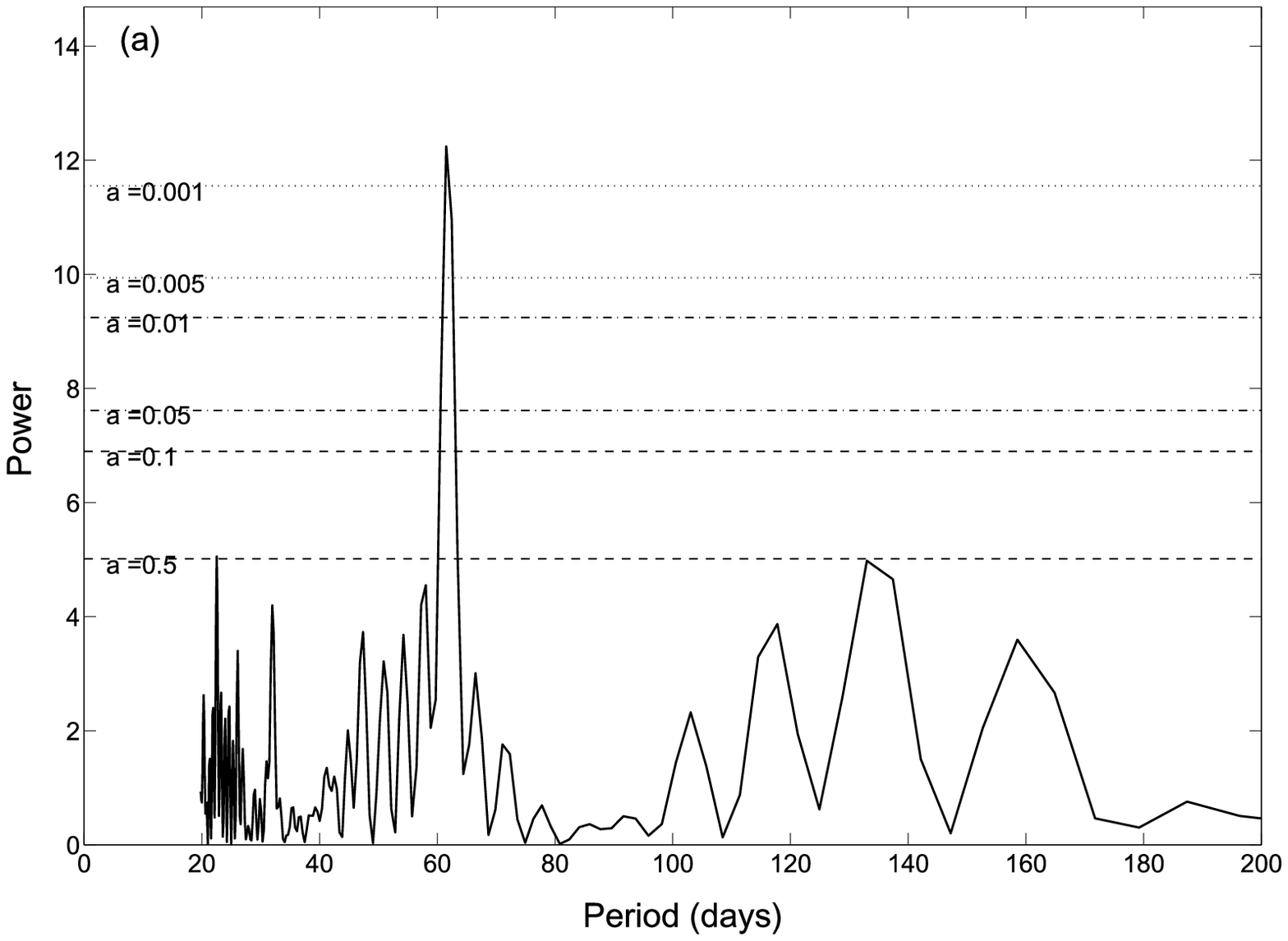}{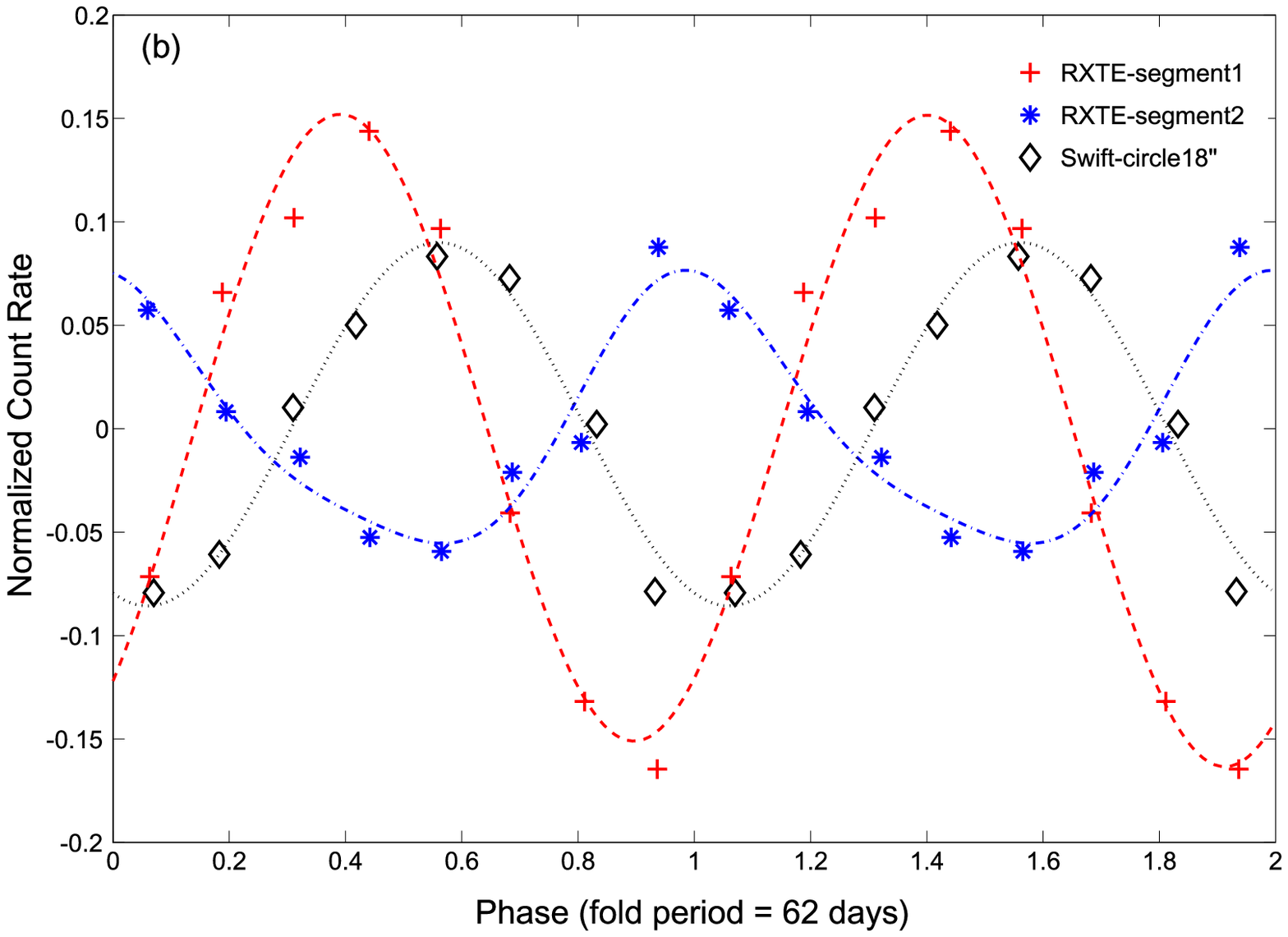}
\caption{
(a) L-S periodogram of a \Swift/XRT circular region centered at X-1 with a
radius of 18\arcsec.  A peak appears at $61.8\pm2.9$ days.  Dashed, horizontal
lines denote the fap of 0.5, 0.1, 0.05, 0.01, 0.005, and 0.001 (or confidence
levels of 50\%, 90\%, 95\%, 99\%, 99.5\%, and 99.9\%), based on the Gaussian noise
background.  (b) Normalized phase light curves of \Swift, \RXTEk segment 1, and
\RXTEk segment 2  folded at 62 days.  Each point is the averaged count rate of
the observations falling within the given phase bin, and the error bar is shown
as the standard deviation divided by the square root of the number of points in
that bin.  The solid sinusoidal curves are the best fourier fit of the data.  }
\label{f2}
\end{figure*}

Based on the long-term observations, \Swift/XRT data can be used to test the 62-day modulation (KF07) of M82.  Source counts were extracted from a circular
region 18\arcseck in radius centered on X-1, with 73\% of energy encircled
\citep{2005SPIE.5898..360M}.  The background was defined by a circular region
with a radius of 20 pixels (47\arcsec) placed near M82 X-1, but avoiding the
contamination from other point sources.
The 0.3-10 keV background-subtracted light curve, binned per observation, was
used to look for the periodicity of X-1 with the Lomb-Scargle(L-S) algorithm
\citep{1976Ap&SS..39..447L, 1982ApJ...263..835S, Press2011}.
The power of the periodogram was normalized by the total variance of the data
\citep{1986ApJ...302..757H}.
As shown in Figure \ref{f2} (a), the main peak appears at the period of
$61.8\pm2.9$ days, and the error is estimated by calculating the FWHM of the main peak in the periodogram \citep{1981Ap&SS..78..175K}.

The significance of the main peak in periodogram can be given by the false alarm probability (hereafter fap) assuming all of the datapoints were generated by stochastic noises. The fap based on the Gaussian white noise of the main peak is less than 0.001 (dashed horizontal lines in Figure 2(a)), \ie, the confidence level of this period generated by a signal instead of the white noise background is more than 99.9\%.
While significant peaks under the white noise background may have low significance when taking the red noise component into consideration.
In contrast to white noise, whose power is independent of frequency, red noise exhibits
a power-law spectrum in the form of $P(\omega)\propto f^{-\beta}$ ($\beta=0$ for white noise), which can lead to relatively large aperiodic peaks in the power spectrum at low frequencies \citep{2011A&A...533A..61G}.

In actuality, the periodogram (Figure 2(a)) does not show evidence of increasing slope for red noise at long periods, but is rather flat with a best-fit power law index of $-0.32$ \citep{2005A&A...431..391V}.
In spite of this, we estimated the significance of the observed data under a red noise background.
100,000 red noise synthetic light curves with a mean and variance equal to the ones of the actual data were generated \citep{1987Biometrika..74..101, 1995A&A...300..707T}.
Each synthetic light curve has the same sampling time series as the actual data set
and was used to conduct L-S periodograms (see \citealt{2006ApJ...646..174K} for more details).
We found only two cases that power values at a period of $61.8\pm2.9$ days excess the power (12.24)
of the observed highest peak. Namely, the probability that a red noise produces a peak at the given period ranges higher than 12.24 is only $2\times10^{-5}$.
The fap of the highest peak at the inspected frequencies (40-80 days) based on red noise is 0.0012 (confidence of 99.88\%) \citep{2015arXiv150203113C}.
Hence, the 62-day periodicity from \Swift/XRT data is significant enough to be produced by a real periodic signal, and confirms the $\sim 62-$day modulation of M82 discovered by KF07 with \RXTE/PCA.

The phase shift of the 62-day period was reported by PS13 utilizing \RXTE/PCA
observations.  Following the work in PS13, we folded the new \Swift/XRT light
curve (MJD 56022.9-MJD 57053.4) at 62 days to test the stability of the $\sim
62-$day periodicity.
The previous \RXTEk light curves from PS13 consist of two segments, which cover
MJD 53250.6-54225.4 (segment 1, the same data used in KF07) and MJD 54353.8-55195.8 (segment 2), respectively.  Start times of the \Swiftk light curve
and the two \RXTEk light curves were set to be the same (MJD 53250.6).  Then
the three light curves were folded at the period of 62 days (see Figure
\ref{f2}(b)).
The two RXTE light curves indeed reveal a 0.4 phase offset as reported in PS13.
Compared with RXTE light curves, the \Swiftk light curve shows changes of
about 0.2 in phase relative to \RXTEk segment 1, and 0.4 in phase relative to
\RXTEk segment 2.  It confirms the result in PS13 that the 62-day X-ray
periodicity of M82 has phase changes, indicating that it is not stable.

\subsection{Origin of the periodicities in M82}

\begin{figure*}
\plottwo{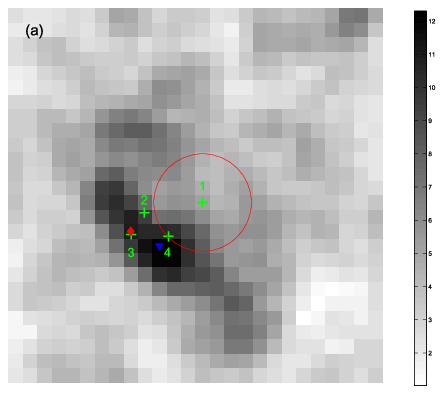}{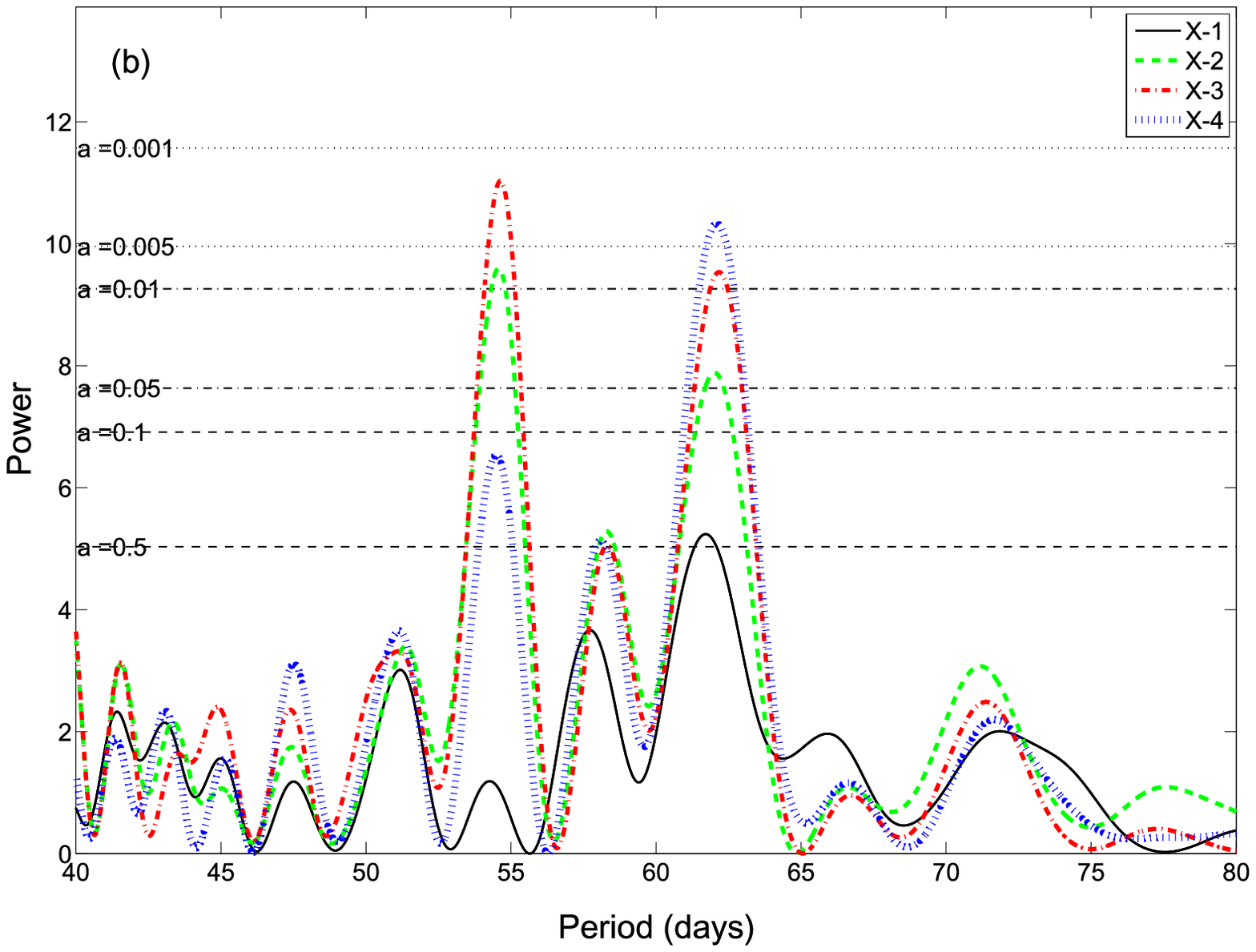}
\caption{
(a) Periodicity map showing the L-S periodogram maximum power for a grid
with steps of 0.5 pixels (i.e., 1\arcsec.18).  The periodograms were computed
with photons from subpixels within a 4\arcseck circular region centered at each
grid point. The red circle is centered on X-1 with a radius of 4\arcsec.  The
red diamond and blue  triangle mark the center of the approximate 55- and 62-
day periodicities respectively. It is clear that there is no significant periodicity
within the 4\arcsec circular region of X-1. (b) L-S periodograms of X-1
(black filled line), X-2 (green line), X-3 (red line), and X-4 (blue line)
extracted from source region of 4\arcseck in radii.  It shows no significant
periodicity in X-1, whereas X-2 and X-3 show two significant periodicities at
$\sim55$ and $\sim62$ days, and X-4 shows the strongest period at 62 days.
The confidence levels of the highest peaks in X-1, X-2, X-3, and X-4 based on red noise are 69\%, 99.6\%, 99.7\%, and 99.8\% respectively.
Dashed, horizontal lines have the same meaning as in Figure2 (a).
}
\label{f3}
\end{figure*}

Since the 18\arcseck radius circular region used in \S2.1 encircles not only
X-1 but many other X-ray sources, it is not clear whether the 62-day periodicity
originates from a certain single source, like X-1, or from summed contributions
of several X-ray sources.
Hence, we used a smaller source region with a radius of 4\arcsec, which can
separate X-1 from other X-ray sources, yet contains 30\% of the energy
encircled in the 18\arcseck region used in \S2.2, to construct the periodicity
map of M82 and analyze the origin of the 62-day modulation.
We divided the region around M82 X-1 into a grid with steps of 0.5 pixel (i.e.,
1\arcsec.18) in X and Y directions, and constructed light curves with photons
from the sub-pixels (as in \S2.1) within a 4\arcseck circular region centered at
each grid point.
L-S periodograms were calculated as in \S2.2, and the power for the strongest
peak was chosen as the value for each grid point to construct the periodicity
map as displayed in Figure \ref{f3}(a).

This periodicity map exhibits two strong periods at $54.6\pm 2.1 $
days and $62.0\pm 3.3$ days (denoted by the red diamond and the blue triangle, respectively, in
Figure \ref{f3}(a)), which are positionally coincident with X-3 and X-4
respectively.  In contrast, there is almost no significant periodicity near X-1.
The individual periodograms of X-1, X-2, X-3, and X-4 with source regions of
4\arcseck in radii are displayed in Figure \ref{f3}(b).
The highest peak in the periodogram of X-1 is at $\sim62.0$ days, but with
much lower confidence levels of 56\% and 69\% for white noise and red noise background respectively.
Except for X-1, the periodograms of X-2, X-3, and X-4 all show the $\sim55-$ day and $\sim62-$ day period, and the significance of the highest peaks in them are larger than 99\%  based on both white noise and red noise background.
This implies that the $\sim55-$ and $\sim62-$ days peaks are generated by real periodic signals with high probability rather than white or red noises,
and the photons from the two signals are
mixed up in the region containing X-2, X-3, and X-4 (dark part in Figure
\ref{f3}(a)), while they still can be distinguished from the counts of X-1.
Thus the 62-day periodicity is most likely associated with the region 4\arcseck southeast of X-1, instead of  X-1 as expected.

\section{DISCUSSION}

By combining the \RXTEk and \Swiftk light curves, we analyzed a longer term
light curve. The analysis confirms phase changes of the 62-day period of M82,
indicating that this period is not stable. A detailed timing analysis of M82 shows
two significant periodicities at 55 and 62 days in a region 4\arcseck southeast
of X-1, which contains X-2, X-3, and X-4. Given the X-ray localization precision of
XRT, the periodicity map can hardly resolve X-2, X-3, and X-4
clearly, but it can easily distinguish the three ULXs from X-1. This
indicates that the 62-day period is not from M82 X-1, but most likely from the
three ULXs. The instability of the 62-day period may be due to the
comprehensive effect of several different periodic X-ray sources, instead of a
precessing accretion disk.

It is not yet clear which sources produce the $\sim55-$ and $\sim62-$ day
periodicities. However, X-4 can be excluded first since it is a luminous young
X-ray supernova remnant candidate \citep{2007ApJ...671..349K, 2011MNRAS.414.1329C},
whose flux cannot fluctuate periodically. X-2 (also
referred to as X42.3+59) is interpreted as a magnetized neutron star system
with an orbital period of 2.5 days \citep{2014Natur.514..202B}. Superorbital
periods of neutron star binaries can range from 25 days to 160 days \citep{1999MNRAS.308..207W}.
The $\sim55$ days may be a superorbital period of X-2, but it may be not valid because the X-ray luminosity of X-2 is several orders of magnitude larger than for the other neutron star binaries \citep{1995PASJ...47..575M, 1997MNRAS.288...43R, 1998ApJ...500L.163O, 2008ApJ...675.1487S}.

For X-3, it is suggested as a background active galactic nucleus because of its
soft excess below 2 keV \citep{2007ApJ...671..349K}. However, soft excess is
also common in an obscured high-mass X-ray binary (HMXB; \eg, IGR J19140+0951 and
IGR J16207-5129; \citealt{2008MNRAS.389..301P, 2009ApJ...694..344T}). X-3 has also
been regarded as an HMXB \citep{2012MNRAS.419.2095M}. Given the
strong star-forming activities in M82, HMXBs are expected to dominate the
bright X-ray sources \citep{2011MNRAS.414.1329C}.  Assuming the isotropic
unabsorbed emission, the maximum 0.5-10 keV luminosity of X-3 is $1.7 \times
10^{40}$ \ergs \citep{2007ApJ...671..349K}, which is approximate to the (maximum-minimum) luminosity modulation of M82 ($\sim 1.6 \times 10^{40}$ \ergs; KF07). Therefore, X-3 is bright enough to dominant the periodicity in M82 if
X-3 is an HMXB with an orbital period of 62 days.

In summary, the limited spatial resolution of XRT is the bottleneck in this
study, but it is already indicating that the 62-day periodicity of M82 may not come
from X-1. To obtain a reliable location of the periodicities, more monitoring
of M82 is needed.

~\\ We thank the anonymous referee for a number of helpful comments. We are grateful for the service of the \Swiftk Data Archive. We thank Song
Wang, Yu Bai, Stephan Justham and Qing Gao \etal \ for useful discussions.  We
would like to specially thank D. J. Pasham for providing \RXTEk reduced data and
helpful discussions. We gratefully acknowledge support for this work from the
National Science Foundation of China through grants NSFC-11333004/11425313/11473036.


\medskip
\begin{thebibliography}{38}
\expandafter\ifx\csname natexlab\endcsname\relax\def\natexlab#1{#1}\fi


\bibitem[Bachetti et al.(2014)]{2014Natur.514..202B} Bachetti, M.,
Harrison, F.~A., Walton, D.~J., et al.\ 2014, \nat, 514, 202



\bibitem[Charisi et al.(2015)]{2015arXiv150203113C} Charisi, M., Bartos,
I., Haiman, Z., Price-Whelan, A.~M.,
\& M{\'a}rka, S.\ 2015, arXiv:1502.03113

\bibitem[Chiang
\& Kong(2011)]{2011MNRAS.414.1329C} Chiang, Y.-K., \& Kong, A.~K.~H.\ 2011, \mnras, 414, 1329

\bibitem[Davies \& Harte (1987)]{1987Biometrika..74..101} Davies, R.B., \& Harte, D.S.\ 1987, Biometrika, 74, 95

\bibitem[Fabbiano(2006)]{2006ARA&A..44..323F} Fabbiano, G.\ 2006, \araa, 44, 323


\bibitem[Freedman et al.(1994)]{1994ApJ...427..628F} Freedman, W.~L.,
Hughes, S.~M., Madore, B.~F., et al.\ 1994, \apj, 427, 628


\bibitem[Gladstone
\& Roberts(2009)]{2009MNRAS.397..124G} Gladstone, J.~C., \& Roberts, T.~P.\ 2009, \mnras, 397, 124

\bibitem[Gruber et
al.(2011)]{2011A&A...533A..61G} Gruber, D., Lachowicz, P., Bissaldi, E., et al.\ 2011, \aap, 533, A61

\bibitem[Horne
\& Baliunas(1986)]{1986ApJ...302..757H} Horne, J.~H., \& Baliunas, S.~L.\ 1986, \apj, 302, 757


\bibitem[Kaaret et al.(2001)]{2001MNRAS.321L..29K} Kaaret, P., Prestwich,
A.~H., Zezas, A., et al.\ 2001, \mnras, 321, L29




\bibitem[Kaaret
\& Feng(2007)]{2007ApJ...669..106K} Kaaret, P., \& Feng, H.\ 2007, \apj, 669, 106


\bibitem[Kaaret et al.(2009)]{2009ApJ...692..653K} Kaaret, P., Feng, H.,
\& Gorski, M.\ 2009, \apj, 692, 653


\bibitem[Kaaret et al.(2006)]{2006ApJ...646..174K} Kaaret, P., Simet,
M.~G., \& Lang, C.~C.\ 2006, \apj, 646, 174


\bibitem[Kong et al.(2007)]{2007ApJ...671..349K} Kong, A.~K.~H., Yang,
Y.~J., Hsieh, P.-Y., Mak, D.~S.~Y., \& Pun, C.~S.~J.\ 2007, \apj, 671, 349


\bibitem[Kovacs(1981)]{1981Ap&SS..78..175K} Kovacs, G.\ 1981, \apss, 78, 175


\bibitem[Liu et al.(2013)]{2013Natur.503..500L} Liu, J.-F., Bregman, J.~N.,
Bai, Y., Justham, S., \& Crowther, P.\ 2013, \nat, 503, 500


\bibitem[Liu
\& Mirabel(2005)]{2005A&A...429.1125L} Liu, Q.~Z., \& Mirabel, I.~F.\ 2005, \aap, 429, 1125


\bibitem[Lomb(1976)]{1976Ap&SS..39..447L} Lomb, N.~R.\ 1976, \apss, 39, 447

\bibitem[Matsuba et al.(1995)]{1995PASJ...47..575M} Matsuba, E., Dotani,
T., Mitsuda, K., et al.\ 1995, \pasj, 47, 575

\bibitem[Matsumoto et al.(2001)]{2001ApJ...547L..25M} Matsumoto, H., Tsuru,
T.~G., Koyama, K., et al.\ 2001, \apjl, 547, L25


\bibitem[Mineo et al.(2012)]{2012MNRAS.419.2095M} Mineo, S., Gilfanov, M.,
\& Sunyaev, R.\ 2012, \mnras, 419, 2095


\bibitem[Moretti et al.(2005)]{2005SPIE.5898..360M} Moretti, A., Campana,
S., Mineo, T., et al.\ 2005, \procspie, 5898, 360


\bibitem[Motch et al.(2014)]{2014Natur.514..198M} Motch, C., Pakull, M.~W.,
Soria, R., Gris{\'e}, F., \& Pietrzy{\'n}ski, G.\ 2014, \nat, 514, 198


\bibitem[Motta et al.(2014)]{2014MNRAS.437.2554M} Motta, S.~E., Belloni,
T.~M., Stella, L., Mu{\~n}oz-Darias, T.,
\& Fender, R.\ 2014, \mnras, 437, 2554

\bibitem[Orlandini et al.(1998)]{1998ApJ...500L.163O} Orlandini, M., Dal
Fiume, D., Frontera, F., et al.\ 1998, \apjl, 500, L163


\bibitem[Pasham
\& Strohmayer(2013)]{2013ApJ...774L..16P} Pasham, D.~R., \& Strohmayer, T.~E.\ 2013, \apjl, 774, L16


\bibitem[Pasham et al.(2014)]{2014Natur.513...74P} Pasham, D.~R.,
Strohmayer, T.~E., \& Mushotzky, R.~F.\ 2014, \nat, 513, 74


\bibitem[Prat et al.(2008)]{2008MNRAS.389..301P} Prat, L., Rodriguez, J.,
Hannikainen, D.~C., \& Shaw, S.~E.\ 2008, \mnras, 389, 301

\bibitem[Press et al.(1992)]{1992nrca.book.....P} Press, W.~H., Teukolsky,
S.~A., Vetterling, W.~T.,
\& Flannery, B.~P.\ 1992, Cambridge: University Press, |c1992, 2nd ed.,



\bibitem[Reynolds et al.(1997)]{1997MNRAS.288...43R} Reynolds, A.~P.,
Quaintrell, H., Still, M.~D., et al.\ 1997, \mnras, 288, 43



\bibitem[Scargle(1982)]{1982ApJ...263..835S} Scargle, J.~D.\ 1982, \apj,
263, 835


\bibitem[Strohmayer
\& Mushotzky(2003)]{2003ApJ...586L..61S} Strohmayer, T.~E., \& Mushotzky, R.~F.\ 2003, \apjl, 586, L61


\bibitem[Suchy et al.(2008)]{2008ApJ...675.1487S} Suchy, S., Pottschmidt,
K., Wilms, J., et al.\ 2008, \apj, 675, 1487


\bibitem[Timmer
\& Koenig(1995)]{1995A&A...300..707T} Timmer, J., \& Koenig, M.\ 1995, \aap, 300, 707


\bibitem[Tomsick et al.(2009)]{2009ApJ...694..344T} Tomsick, J.~A., Chaty,
S., Rodriguez, J., et al.\ 2009, \apj, 694, 344

\bibitem[Vaughan(2005)]{2005A&A...431..391V} Vaughan, S.\ 2005, \aap, 431, 391

\bibitem[Wijers
\& Pringle(1999)]{1999MNRAS.308..207W} Wijers, R.~A.~M.~J., \& Pringle, J.~E.\ 1999, \mnras, 308, 207






\end{thebibliography}
\end{document}